\documentclass[preprint,11pt]{aastex}

\usepackage{graphicx}
\usepackage{amsmath,amssymb}
\usepackage{natbib}
\usepackage{array}

\usepackage{graphicx}
\usepackage{amssymb,amsmath}

\title{Detection of pulsar beams deflected by the black hole in Sgr A$^*$: effects
of black hole spin}

  \author{Sourabh Nampalliwar} 

\affil{ Department of Physics \& Astronomy and Center for Gravitational
  Wave Astronomy, University of Texas at Brownsville, Brownsville,
  Texas 78520}

\author{Richard H.~Price, Teviet
    Creighton and Fredrick A. Jenet }

  \affil{Department of Physics \&
    Astronomy, Center for Gravitational Wave Astronomy and  
Center for Advanced Radio Astronomy, University of Texas at Brownsville, Brownsville, Texas 78520}

\begin{document}

\pagebreak
\maketitle

\begin{abstract}
Some Galactic models predict a significant population of radio pulsars
close to the our galactic center. Beams from these pulsars could get
strongly deflected by the supermassive black hole (SMBH) believed to
reside at the galactic center and reach the Earth. Earlier work
assuming a Schwarzschild SMBH gave marginal chances of observing this
exotic phenomenon with current telescopes and good chances with future
telescopes. Here we calculate the odds of observability for a rotating
SMBH. We find that the estimates of observation are not affected by
the SMBH spin, but a pulsar timing analysis of deflected pulses might
be able to provide an estimate of the spin of the central black hole.
\end{abstract}

\section{Introduction}
\label{sec:intro}
It is believed that the center of essentially every large galaxy, including
our own, contains a supermassive black hole (SMBH). Sagittarius
A*, first discovered as a bright radio source,
is expected to host the SMBH at the center of the Milky
Way. Observations of the region in the
infrared \citep{genzel2003,gillessen2009},
radio \citep{doeleman2008,broderick2009},
x-ray \citep{baganoff2003,porquet2003} and
optical \citep{ghez2008}, have revealed
properties of the central object such as mass and 
accretion flow. Alternates to an SMBH, such as a dense cluster of dark stellar
objects or a ball of massive, degenerate
fermions \citep{schodel2002} have largely been ruled out. These observations also indicate an
initial mass function (IMF) near the SMBH strongly tilted towards
massive
stars \citep{nayakshin2005,maness2007,levin2006}. This
top-heavy IMF suggests a large population of neutron stars orbiting
the SMBH with estimates placing $\sim$10000 of them within 1 pc of Sgr
A* \citep{muno2005} and 1000 within 0.01
pc \citep{pfahl2004}. Beams from these pulsars could traverse
the strong field region of the SMBH, be  deflected and reach the
Earth. Using the precision of pulsar timing, we might be able to
detect these deflected beams and from them infer properties of the strong
field region through which the beams have passed.

Previous work in this direction with the assumption of a nonspinning SMBH has been reported in
Wang et al. \citep{wang2009a,wang2009b} and Stovall
et al. \citep{stovall2011} (hereafter Paper I, II and III
respectively). In Paper I, strong field effects on the intensity and time
of arrivals (TOA) of pulsar beams were analyzed in the case
that the pulsar beam is confined to the plane of the pulsar orbit. Paper II extended the formalism to general
orientations. In Paper III, an
estimate was given  of the probability of observing these deflected pulsar beams.
For the pulsar luminosity distribution calculated in the ATNF Pulsar Catalog\footnote{\url{http://www.atnf.csiro.au/research/pulsar/psrcat/}}, it was found that there is a marginal probability of detection 
with current instruments, like the Parkes and Green Bank Telescope, and future instruments, in
particular FAST and SKA, provide a reasonable chance of detection. For a more conservative distribution \citep{faucher2006}, chances of detection with existing telescopes are weak, and we must await more sensitive instruments like SKA. 

Paper III was based on the the model of Paper II, and hence embodied
the assumption of a nonspinning SMBH. The Sgr A* SMBH is certainly
expected to be rotating, though the rate of rotation is indeterminate,
with predictions for the spin rate ranging from very
slow \citep{liu2002} to moderate \citep{genzel2003}. In principle, spin
effects could have a dramatic effect on beam bending and hence on the
probability of the detection of deflected beams. The chief motivation
of the current paper is to analyze a pulsar-Kerr SMBH system to see if
the detection probability estimates obtained for a nonrotating SMBH are
significantly different, and especially whether they might improve
the prospects for near-term detection. A second purpose of this paper 
is to check whether the SMBH spin would have a strong enough effect
on pulsar timing that spin information could be extracted from the 
arrival times of the pulses in a deflected beam. 

The paper is organized as follows: in Section~\ref{sec:review}, we
give an overview of the previous work and redefine terms that will be useful for
purposes of generalization and comparison. We compare deflection of beams by a rotating SMBH and that by a nonrotating SMBH in Section~\ref{sec:schvskerr}. Section~\ref{sec:probability}
describes the geometry of the system: beams emitted from a pulsar
revolving around a spinning SMBH. This Section also outlines the scheme for
calculating probabilities for a
detection. Section~\ref{sec:appearance} compares the range of
detection of deflected beams in a pulsar/spinning-SMBH system to that
for a pulsar/Schwarzschild-SMBH system. Section~\ref{sec:timing} compares the
pattern of pulse times of arrival in the two scenarios. We conclude in
Section~\ref{sec:conclusion}. Throughout the paper, we will use the
natural units ($c = 1, G = 1$) and describe distances and time in terms
of the black hole mass $M$.

\section{Review of the Schwarzschild case}
\label{sec:review}
As observed in Paper III, for a pulsar revolving around a
Schwarzschild SMBH, there is a `region of detectability' - a range of
Earth angular positions on the pulsar sky at which both direct and
deflected beams are detectable. We start by reviewing the determination of
this region.

\begin{figure}[ht]
  \begin{center}
    \includegraphics[scale=0.60]{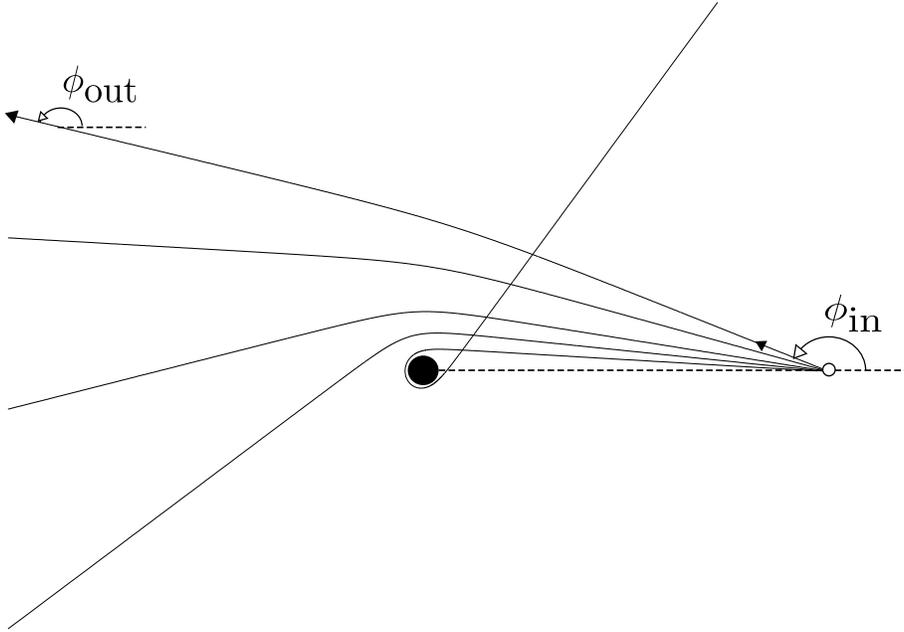}
  \end{center}
  \caption{Trajectories of photons passing through the strong field
    region of a Schwarzschild black hole, emitted by a pulsar far away
    from the strong field region. Angles $\phi_{\textrm{in}}$ and
    $\phi_{\textrm{out}}$ are illustrated for one of the
    trajectories. This figure also serves for
photons emitted in the equatorial plane of a
    spinning black hole.}
  \label{fig:bending}
\end{figure}

Figure~\ref{fig:bending} sketches the trajectory of photons outside a
Schwarzschild black hole. Also illustrated is the set of angles we
use to describe a particular trajectory. Here $\phi_{\textrm{in}}$
corresponds to the angle at which the photon is emitted relative to
the radially outward direction at the point of emission in a spherical
coordinate system centered at the SMBH; $\phi_{\textrm{out}}$
corresponds to the angle measured relative to the same radial
direction far away from the strong field region of the SMBH. Paper I
relates these two angles by a function $F$:
\begin{equation}
  \label{eq:defF}
  \phi_{\textrm{out}} = F(\phi_{\textrm{in}};r_0)
\end{equation}
where $r_0$ is the pulsar-SMBH distance at the time of emission. In
the absence of deflection, $\phi_{\textrm{out}}$ will be equal to
$\phi_{\textrm{in}}$.

\begin{figure}[ht]  
  \begin{center}
    \includegraphics[scale=0.60]{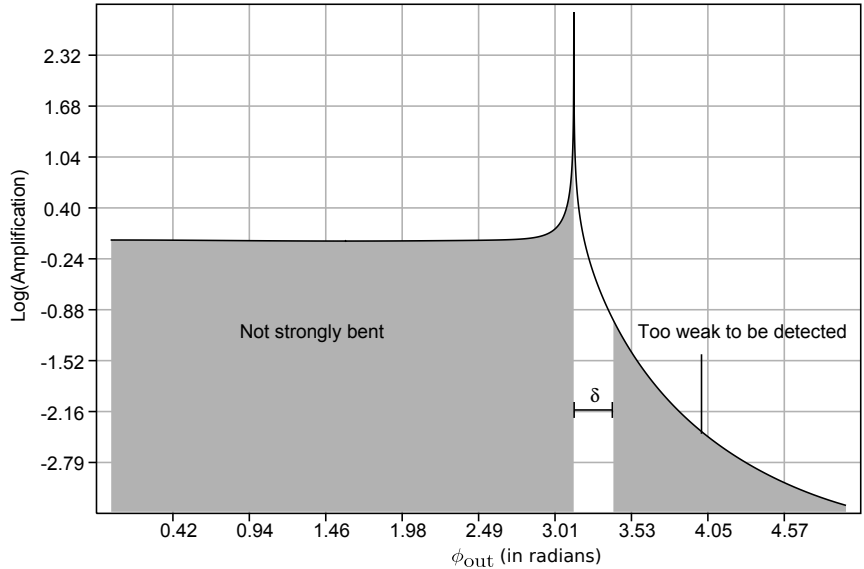}
  \end{center}
  \caption{Intensity vs $\phi_{\textrm{out}}$, for a pulsar at $100M$ from
    a Schwarzschild black hole. The shaded regions represent beams that are
    either weakly bent or too strongly attenuated.}
  \label{fig:attvsphiout}
\end{figure}

Paper III notes that the intensity of a deflected beam will be
modified. There is a narrow range of $\phi_{\textrm{out}}$ around
$\phi_{\textrm{out}}=\phi_{\textrm{out}|\textrm{peak}}= \pi$ for which
there is dramatic amplification corresponding to strong lensing, but
in general strongly bent beams will be
attenuated. Figure~\ref{fig:attvsphiout} shows the relationship of
$\phi_{\textrm{out}}$ and attenuation.  For
$\phi_{\textrm{out}}\lesssim\phi_{\textrm{out}|\textrm{peak}}$ the
bending is not significant; a deflected beam will not be
distinguishable from a direct beam. For
$\phi_{\textrm{out}}\gtrsim\phi_{\textrm{out}|\textrm{peak}}$, beams that
are very strongly deflected are too attenuated to be
detectable. Therefore, we have a $\phi_{\textrm{out}|\textrm{max}}$, a
cutoff for $\phi_{\textrm{out}}$ beyond which attenuation reduces beam
intensity below telescope sensitivity. We define $\delta =
\phi_{\textrm{out}|\textrm{max}} - \phi_{\textrm{out}|\textrm{peak}}$
as the range of $\phi_{\textrm{out}}$ for which the deflection is
significant, but the deflected beam is still detectable. If the Earth
is within this range, the strongly deflected beam will be detectable.

The determination of the 
probability of detection of a deflected 
beam then involves two steps: calculating the probability of the
pulsar beams being deflected to the region of detectibility, and
calculating how often the Earth will lie in this region. This was
achieved in Paper III for nonrotating black holes and it was found
that with current pulsar population models near the Galactic center,
there is a good chance of detection by a multi-year program with
future telescopes. Our primary aim in the current paper is to 
find whether these estimates are significantly changed by a 
consideration of the spin of the SMBH in  Sgr A*.
The first step will be to find a version of Figure~\ref{fig:attvsphiout}
that applies to a spinning SMBH.

\section{Expected effects of SMBH spin}
\label{sec:schvskerr}

\begin{figure}[ht]
  \begin{center}
    \includegraphics[scale=0.50]{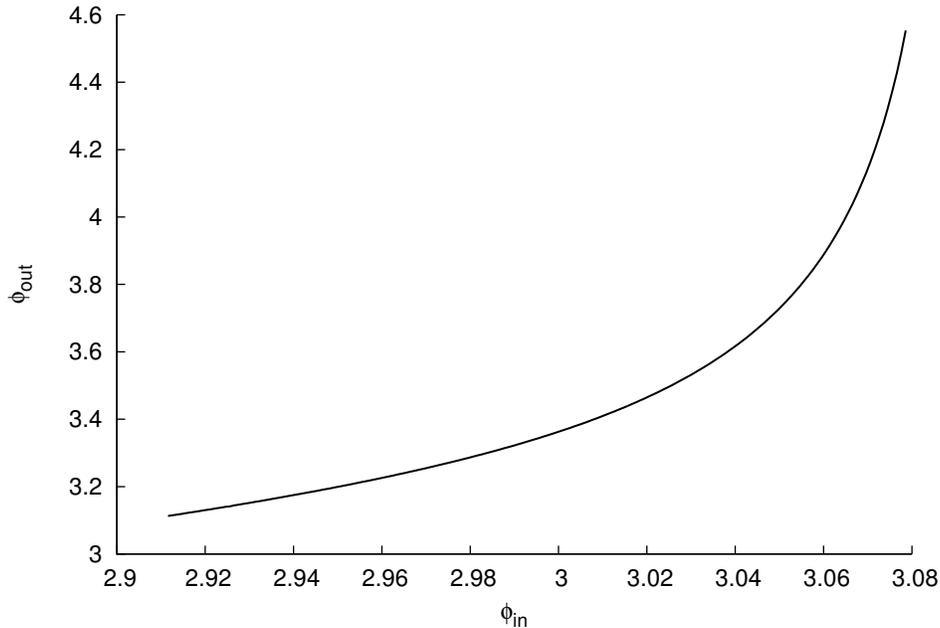}
  \end{center}
  \caption{ The relation of $\phi_{\textrm{out}}$ and
    $\phi_{\textrm{in}}$ for photons emitted at $r_0$ = 100M from a
    Schwarzschild black hole.}
  \label{fig:phioutvsphin}
\end{figure}

The nature of the spin effects on the $F$ function of
Equation~\eqref{eq:defF} are clearest for photon trajectories in the
equatorial plane. Figure~\ref{fig:bending} shows sketches of such
trajectories in either the Kerr geometry or the Schwarzschild geometry
(for which any plane can be considered the equatorial plane).
The deflection is minimum for photons emitted in the radially outward
direction (i.e., for $\phi_{\textrm{in}}$ = 0). As $\phi_{\textrm{in}}$ increases,
photons get closer and closer to the black hole and deflection
increases, as is evident in Figure~\ref{fig:bending}. All photons straying too close, fall into the black hole. We thus identify an orbit around the black hole with the characteristic that the number of orbits of a photon traveling along this orbit grows without bound, so that $\phi_{\textrm{out}}$ effectively approaches infinity. We use the term \emph{unstable circular orbit} (UCO) to refer to this orbit and denote the emission angle of a photon traversing this critical trajectory as $\phi_{\textrm{in}|\infty}$. For a Schwarzschild black hole,
Figure~\ref{fig:phioutvsphin} plots the relationship between
$\phi_{\textrm{out}}$ and $\phi_{\textrm{in}}$ up to
$\phi_{\textrm{in}}= \phi_{\textrm{in}|\infty} $. For larger
values of $\phi_{\textrm{in}}$ there is no $\phi_{\textrm{out}}$
defined since these trajectories end inside the black hole.

For photons orbiting a Schwarzschild hole, it is known that the UCO lies at $r = 3M$. For those orbiting in the equatorial plane of a spinning hole, it turns out such an orbit can be obtained. We observe that for prograde photons, the size of UCO decreases as the black hole spin increases, and for retrograde photons, the size of UCO increases with the black hole spin. A smaller UCO
implies $\phi_{\textrm{in}|\infty}$ for a prograde beam outside
a spinning black hole will be larger than for a nonspinning black
hole. Similarly, a larger UCO for retrograde beams means its
$\phi_{\textrm{in}|\infty}$ will be smaller than that of the
Schwarzschild case.

\section{Calculation of Probability}
\label{sec:probability}
\begin{figure}[ht]
  \begin{center}
    \includegraphics[scale=0.30]{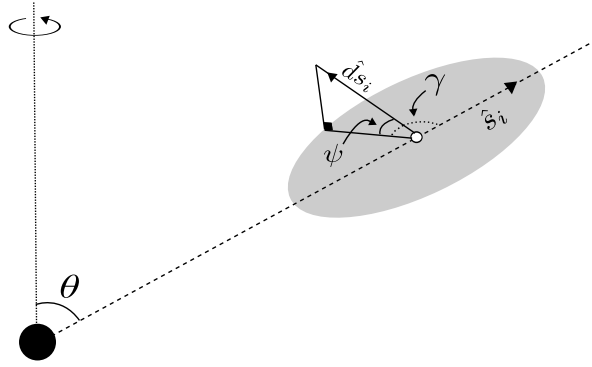}
  \end{center}
  \caption{Geometry of a pulsar-Kerr system. The shaded ellipse is a
    disk on the plane containing the pulsar and Kerr SMBH.}
  \label{fig:angledefKerr}
\end{figure}
The basic geometry outside a Kerr SMBH is illustrated in
Figure~\ref{fig:angledefKerr}. We will use  spherical coordinates
 $\{r,\theta,\phi\}$ centered at the black hole. In the strong field region, the 
region of beam bending,
these will be the spatial coordinates of the Boyer-Lindquist
system. Since the pulsars of interest are all very far from 
the SMBH compared to distance $M$, we can consider the spatial geometry 
to be flat in the neighborhood of the pulsar.  In the equations below, the position of the
pulsar is specified by r$_i$, and $\theta_i$ is the angle the pulsar makes
with the black hole spin axis at the time of emission. The angles $\gamma$ (the
planar angle) and $\psi$ (the axial angle), describing the direction
of photon emission are defined as

\begin{subequations}
  \label{eq:defgammapsi}
    \begin{align}
      \cos{\gamma} & = \hat{s_i}\cdot\hat{p}(\vec{ds_i}) \\    
      \cos{\psi} & = \hat{ds_i}\cdot\hat{p}(\vec{ds_i})
    \end{align}
  \end{subequations}
where $\hat{s_i}$ is the unit vector in the radially outward direction
at the pulsar and $\hat{ds_i}$ is a unit vector denoting the initial
direction of the emitted photon.  The vector $\hat{p}(\vec{ds_i})$
denotes the projection of $\hat{ds_i}$ in the orbital plane. 
Four integrals of 
the equations of motion of a photon in the Kerr geometry can be found.
In Boyer-Lindquist coordinates, these are
 \citep{mtw}
\begin{subequations}
  \label{eq:geodesic}
    \begin{gather}
      \rho^2 d\theta / d\lambda = \sqrt{\Theta}, \\
      \rho^2 dr/d\lambda = \sqrt{R}, \\
      \rho^2 d\phi/d\lambda = -(aE - L_z/\sin^2{\theta}) + (a/\Delta)P, \\
      \rho^2 dt/d\lambda = -a(aE\sin^2{\theta} - L_z) + (r^2 + a^2)\Delta^{-1}P\,,
    \end{gather}
  \end{subequations}
where
\begin{subequations}
  \label{eq:geofunc}
    \begin{gather}
      \Delta = r^2 - 2Mr + a^2, \\
      \rho^2 = r^2 + a^2\cos^2{\theta}, \\
      \Theta = {\cal Q} - \cos^2{\theta}[-(aE)^2 + L_z^2/\sin^2{\theta}], \\
      P = E(r^2 + a^2) - L_za, \\
      R = P^2 - \Delta[(L_z-aE)^2 + {\cal Q}].
    \end{gather}
  \end{subequations}
Here, $r,\theta$ and $\phi$ are the (Boyer-Lindquist) coordinates of
the photon at different values of affine parameter $\lambda$, and $a=J/M$
is the spin parameter of the hole, where $J$ is the angular
momentum of the SMBH. The parameters $L_{z}$, $E$ and ${\cal Q}$
(Carter's constant) are the constants of motion.

Our approach will be to initialize $a$, $r_i$ and $\theta_i$.  We will
then use a public FORTRAN code developed by Dexter and
Agol \citep{dexter2009} to evaluate the trajectory of a photon through
the strong field region. Along with $a$, $r_i$ and $\theta_i$, the
code takes as input the three constants of motion and the final radial
coordinate, which is chosen  large enough that photons undergo
insignificant deflection
for a larger radii. The conversion from our
geometric parameters to the constants of motion is outlined in the
Appendix.

 The code generates final polar coordinates of the photon as
 output. To calculate the attenuation for any given beam, we construct
 a circular grid of photons closely arrayed around a central photon.
 Using the Dexter \& Agol code we then calculate the final positions
 of the central photon and of each photon on the grid. The ratio of
 the initial and final cross sectional areas of the grid of photons
 then gives the attenuation, indicating the change in intensity of the
 beam.  Finally we check that the initial grid is small enough that
 changing to a smaller cross section does not give a substantially
 different result for the attenuation ratio.


\section{Results: Observability of beams}
\label{sec:appearance}
\begin{figure}[ht]
  \begin{center}
    \includegraphics[scale=0.50]{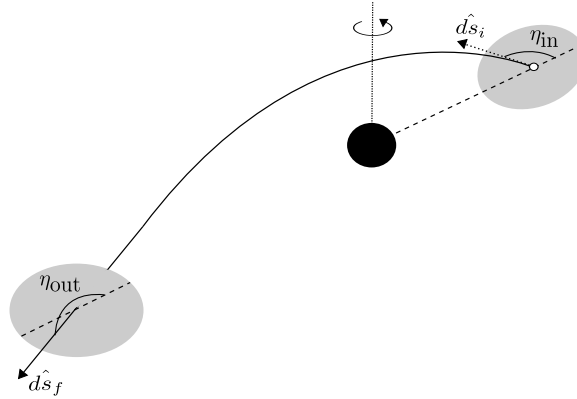}
  \end{center}
  \caption{Definitions of $\eta_{\textrm{in}}$ and $\eta_{\textrm{out}}$ for a photon trajectory around a spinning black hole.}
  \label{fig:etadef}
\end{figure}

To aid us in comparing pulsar-Kerr black hole system with the
pulsar-Schwarzschild black hole system we introduce another set of
angles $\eta_{in}$ and $\eta_{out}$, illustrated in
Figure~\ref{fig:etadef} and defined by
\begin{subequations}
  \label{eq:etadef}
  \begin{align}
    \cos{\eta_{\textrm{in}}} & = \hat{s_{i}}\cdot\hat{ds_{i}} = \cos{\gamma} \cos{\psi} \\
    \cos{\eta_{\textrm{out}}} & = \hat{s_{i}}\cdot\hat{ds_{f}}
  \end{align}
\end{subequations} 
where $\hat{s_{i}}$ and $\hat{ds_{i}}$ were defined in
Section~\ref{sec:probability}, and where $\hat{ds_{f}}$ is the spatial
unit vector in the direction of the final motion of the photon far
from the the pulsar-SMBH system. The interpretation of $\eta_{\textrm{in}}$ in terms of
geometric parameters $\gamma$ and $\psi$ comes from their definitions
in Figure~\ref{fig:angledefKerr} and
Equation~\eqref{eq:defgammapsi}. Far  from the strong
field region, $\hat{ds_{f}}$ and $\hat{s_{f}}$ point in the same
direction. Then, using Equation~\eqref{eq:defs} from the Appendix for $\hat{s}_i$ and $\hat{s}_f$ and setting $\phi_i$=0 for simplicity, the value of 
$\eta_{\textrm{out}}$ can be shown to be given by
\begin{equation}
  \label{eq:etaout}
    \cos{\eta_{\textrm{out}}} = \cos{\theta_{i}} \cos{\theta_{f}} + \sin{\theta_{i}} \sin{\theta_{f}} \cos{\phi_{f}}\ .
\end{equation}

\begin{figure}[ht]
  \begin{center}
    \includegraphics[scale=0.50]{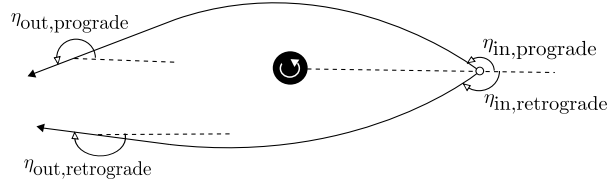}
  \end{center}
  \caption{Prograde and retrograde $\eta$  in the 
equatorial plane of a Kerr black hole.}
  \label{fig:etaproretro}
\end{figure}

It is expected that effects of spin of the black hole on the
deflection of photons will be most prominent in the equatorial
plane. As a first step in our investigation of spin effects, we will
therefore compute attenuation curves for equatorial Kerr photons
similar to the curves in Equation \eqref{fig:attvsphiout}. We will
assume the pulsar and the photon trajectories lie in the equatorial
plane of the SMBH. (In terms of our geometrical parameters, this
corresponds to $\theta_i$ = $\pi$/2 and $\psi$ = 0). Figure
\ref{fig:etaproretro} shows the $\eta$ angles for this simple
scenario. The definition of the $\eta$'s in this scenario is similar to the $\phi$'s
from the Schwarzschild case (Figure~\ref{fig:bending}), and the $\eta$'s approach
the $\phi$'s as the black hole spin goes to zero.

\begin{figure}[ht]
  \begin{center}
    \includegraphics[scale=0.6]{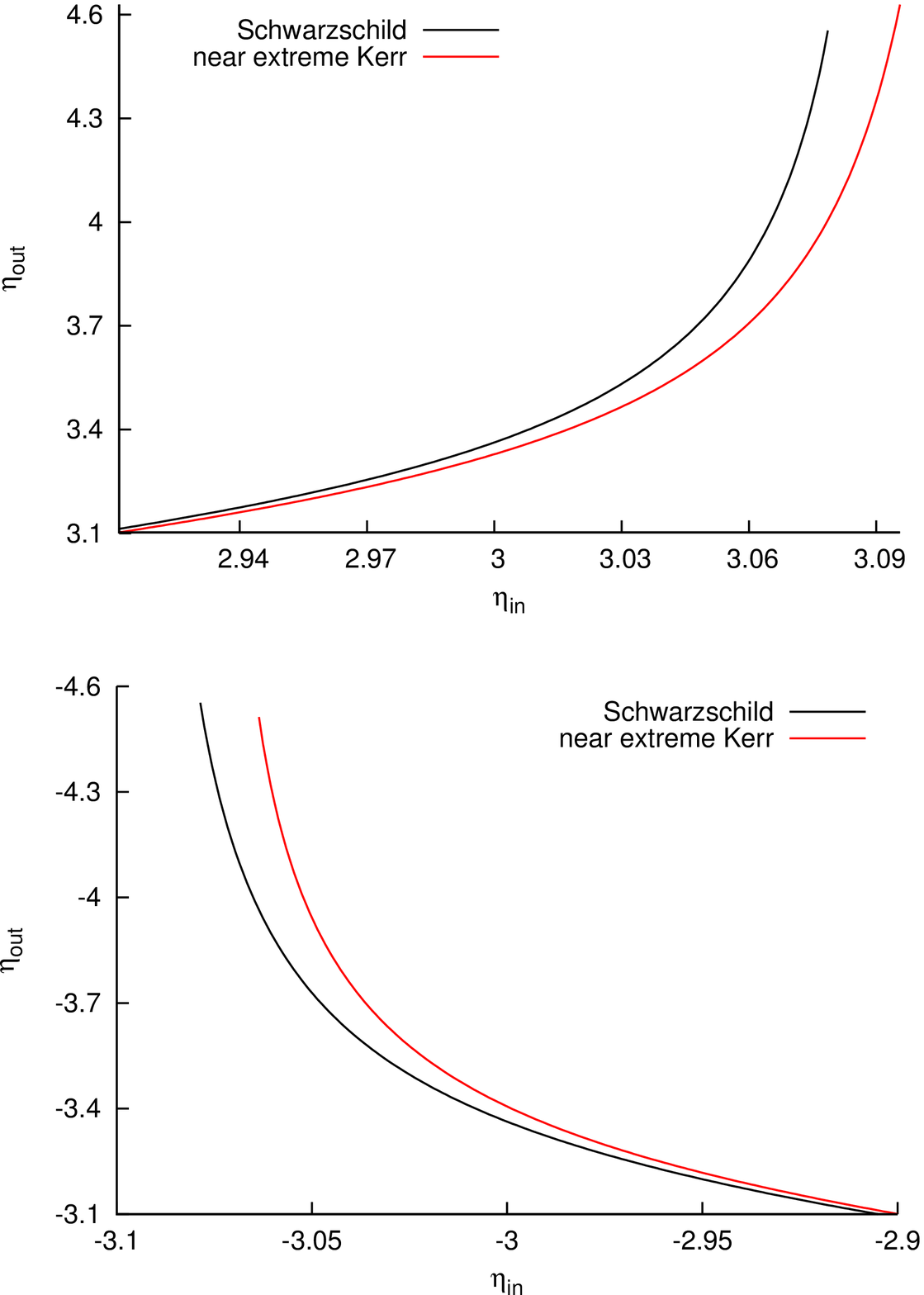}
  \end{center}
  \caption{The relationship of $\eta_{\textrm{out}}$ and
    $\eta_{\textrm{in}}$. The top figure shows prograde beams in the
    equatorial plane of an $a = 0.99$ SMBH (red) and Schwarzschild SMBH
    (black); the bottom figure shows the same comparison for retrograde beams.}
  \label{fig:etaoutvsin}
\end{figure}

 Figure~\ref{fig:etaoutvsin} plots $\eta_{\textrm{out}}$ as a function
 of $\eta_{\textrm{in}}$ for both prograde and retrograde beams
 emitted by a pulsar at a distance of 100\,$M$ from an extreme ($a$ =
 0.99) Kerr SMBH and compares them with the curves for a nonrotating
 SMBH.  There is a late divergence of $\eta_{\textrm{out}}$ for
 prograde photons in the near extreme Kerr SMBH case
 compared to the nonrotating case. This implies a larger $\eta_{\textrm{in}|\infty}$ for
 prograde Kerr beams than prograde Schwarzschild beams. Similarly, the
 early divergence of $\eta_{\textrm{out}}$ for retrograde beams
 orbiting the near extreme Kerr SMBH implies a smaller $\eta_{\textrm{in}|\infty}$ for
 retrograde Kerr beams than retrograde Schwarzschild beams, as was observed theoretically in Section~\ref{sec:schvskerr}.

\begin{figure}[ht]
  \begin{center}
    \includegraphics[scale=0.6]{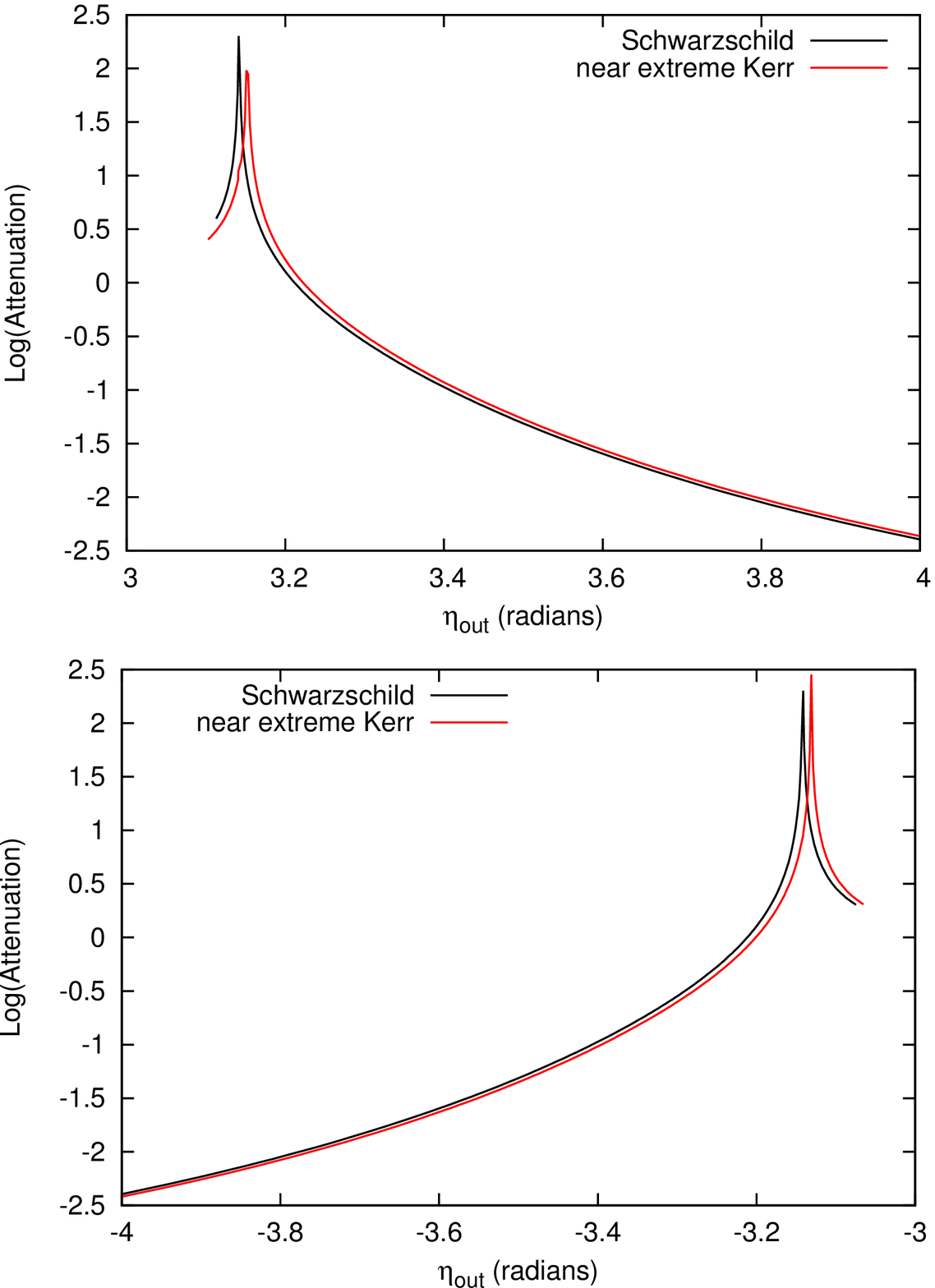}
  \end{center}
  \caption{Attentuation of pulse intensity as a function of
    $\eta_{\textrm{out}}$ for the same models as in
    Figure~\ref{fig:etaoutvsin}. The top figure shows the comparison 
of the Schwarzschild curve with the prograde case, the bottom with 
the retrograde case.}
  \label{fig:attvseta}
\end{figure}
Attenuation curves are plotted in Figure~\ref{fig:attvseta} for the
same model as in Figure~\ref{fig:etaoutvsin}.  We notice that spin
introduces a very small shift in the attenuation curve. The all-important angle $\delta$
increases by only 3$\%$ for prograde beams in the near extreme Kerr case
compared to the Schwarzschild case, and there is  almost exactly the same decrease
in $\delta$ for the retrograde case. We also notice that as
deflection increases the curves are more in agreement,
$\eta_{\textrm{out}}$ at the maximum acceptable attenuation being the same for prograde and
retrograde beams in both cases, respectively. Our results show that
the deviations get smaller as we go out of the equatorial plane
(i.e. $\theta$ $\neq$ $\pi$/2 or/and $\psi$ $\neq$ 0). For larger,
more relevant pulsar-SMBH distances, the differences become
negligible. Thus, spin induces minimal effect on the attenuation curve
of deflected beams, implying that the range of observability of
deflected beams calculated for the pulsar-Schwarzschild case remains
invariant for a pulsar-near extreme Kerr case. The probabilities
established in Paper III of an observation with the assumption that
the Sgr A* black hole is not rotating are therefore unchanged even if
the black hole is spinning very rapidly.

\section{Results: Timing of Pulses}
\label{sec:timing}

The importance of beam detection lies in the potential to use
deflected beams as a probe of the properties of strongly curved
spacetime. For such a probe to be useful it should be sensitive to
SMBH spin.  We have seen that SMBH spin has a negligible effect on
beam detectability. It is therefore important to ask whether the
effect of spin on pulsar beams is so slight that pulsar timing cannot
be used to detect and measure spin.

We aim here only for an order of magnitude demonstration, so we make
several simplifying assumptions. First, we assume that the pulsar is
in a circular orbit around the SMBH, and that the pulsar beam is
confined to the equatorial plane. The timing analysis of TOAs in this
case has been thoroughly analyzed in Paper I for a nonspinning
SMBH. In that analysis it was shown that there are two relativistic
effects on pulsar TOAs. First there is the timing issue directly
connected to the deflection. If the pulsar spins (relative to distant
stars) at $\omega$, then beams that will arrive at Earth without
strong deflection are pulsed at frequency $\omega$. Strongly
deflected beams, on the other hand must be aimed approximately toward
the SMBH, and hence, for pulsar orbital angular frequency $\Omega$,
are emitted approximately at frequency $\omega\pm\Omega$. The spin of
the SMBH will have a small effect on this frequency, since it will
affect the precise near-SMBH ``target'' towards which the
to-be-deflected beam must be aimed.
For simplicity we will ignore this small geometric effect, and will
consider only the second relativistic effect on TOAs, the effect on
the time of travel for a photon in a deflected beam. 

Since the Earth is nearly stationary, a pulse emitted from a
particular orbital position of the pulsar, and received at the Earth,
has a unique $\eta_{\textrm{out}}$: the angle of the final direction
of the beam with respect to the radially outward direction through the
pulsar at the time of emission in a spherical coordinate system
centered at the Galactic center (defined in
Equation~\eqref{eq:etadef}, and pictured in
Figure~\ref{fig:etadef}). For $\cal T$, the pulse arrival time,
$d{\cal T}/d\eta_{\textrm{out}}$ is then a measure of variation in
arrival times as the pulsar travels through its orbit. For a circular
orbit, the radially outward direction through the pulsar rotates at a
fixed rate, the angular velocity $\Omega$ of the pulsar orbit. In the
absence of strong gravity effects, $d{\cal T}/d\eta_{\textrm{out}}$
would then be a constant. Any deviation from a constant value is
therefore a direct measure of strong gravity effects, and we use this
rate to investigate the sensitivity of TOAs to SMBH spin.

The change in coordinate time as a
pulse traverses the SMBH and reaches the Earth is given by
\begin{equation}
  \label{eq:coordinatetime}
  \mathcal{T}\mathnormal{(\eta_{\textrm{out}};R) =
    \mathrm{2}\int^{r_{\textrm{0}}}_{r_{\textrm{min}}} J(r)dr +
    \int^{R}_{r_{\textrm{0}}} J(r)dr}
\end{equation}
where $r_0$ is the pulsar-SMBH distance, $R$ is the distance between
the SMBH and the Earth and
\begin{equation}
  \label{eq:defineJr}
  J(r) = \frac{dt}{dr} = \frac{P(r^2 + a^2)\Delta^{-1} + aL_z}{\sqrt{R}}
\end{equation}
is obtained from the equations of motion of a photon in the Kerr
geometry (Equations~\eqref{eq:geodesic} and~\eqref{eq:geofunc}). Since
the Earth is very far away from the pulsar-SMBH system, we use 
Equation (\ref{eq:coordinatetime})
to find the value of ${\cal T}$ to
a distance $R$ much larger than $M$, where relativistic effects are negligible.

To compare the prograde and the retrograde pulses for a possible spin
signature, we start with the absolute value of $d{\cal  T}/d\eta_{\textrm{out}}$
as a measure of the variation in the
TOA. We analyze $d{\cal T}/d\eta_{\textrm{out}}$ as a function of the
absolute value of $\eta_{\textrm{out}}$, an indicator of the orbital
position of the pulsar.  If the SMBH is not rotating, the prograde and
the retrograde beams will suffer identical deflection and we find that
this Schwarzschild value
$\left.d{\cal T}/d\eta_{\textrm{out}}\right|_{S}$ is, of course, 
the same in the 
prograde and the retrograde
cases for any $\eta_{\textrm{out}}$. If the SMBH is
rotating, this symmetry is broken; there will be different values of
$d{\cal T}/d\eta_{\textrm{out}}$ for prograde and retrograde orbits.
We use the absolute value of 
$d{\cal T}/d\eta_{\textrm{out}}
-\left.d{\cal T}/d\eta_{\textrm{out}}\right|_{S}
$
as a measure of the spin detectibility.

This measure, and all measures of deflection, depend on the
closest approach of the pulsar beam to the SMBH, and that
closest-approach distance determines the attenuation of the TOAs. We
choose, therefore, to compare the prograde and retrograde values of
$d{\cal T}/d\eta_{\textrm{out}}$ for fixed, relevant values of the
attenuation of the pulsar beam. Table \ref{tab:toadeviationat50} gives
the result for 50\% attenuation, i.e., for significant deflection, 
and for 90\% attenuation, i.e., for the limiting deflection, the deflection
beyond which attenuation is assumed to preclude observation.
The tabulated values are absolute values of 
$d{\cal T}/d\eta_{\textrm{out}} - \left.d{\cal T}/d\eta_{\textrm{out}}\right|_{S}$. This value is the same
for the prograde and retrograde cases. In
the two cases the signs are opposite, so the
total prograde/retrograde difference is twice the values tabulated. Several
properties of spin effects are immediately clear from the tables: 
the spin-induced effect, at any attentuation, is proportional
to the SMBH spin;  the effect depends very weakly 
on the pulsar-SMBH distance; the effect is stronger at the stronger deflection 
of the 90\% deflection, but the sensitivity to deflection is not strong. 
  \begin{table}[h!]
    \begin{center}
    \begin{tabular}{| c || c | c | c| }
      \hline
      50\% atten.&\multicolumn{3}{r|}{pulsar-SMBH distance} \\
    \hline
    SMBH spin & 10$^2M$ & 10$^3M$ & 10$^4M$ \\
    \hline
    0.30 & 0.023 & 0.020 & 0.019 \\
    0.60 & 0.046 & 0.040 & 0.038 \\
    0.99 & 0.075 & 0.067 & 0.063 \\ \hline
  \end{tabular}
{\ }
    \begin{tabular}{| c || c | c | c| }
      \hline
      90\% atten.&\multicolumn{3}{r|}{pulsar-SMBH distance} \\
    \hline
    SMBH spin & 10$^2M$ & 10$^3M$ & 10$^4M$ \\
    \hline
    0.30 & 0.028 & 0.025 & 0.024 \\
    0.60 & 0.056 & 0.049 & 0.048 \\
    0.99 & 0.093 & 0.081 & 0.079 \\ \hline
  \end{tabular}

  \end{center}
\caption{\label{tab:toadeviationat50} 
The value of 
$d{\cal T}/d\eta_{\textrm{out}}
-\left.d{\cal T}/d\eta_{\textrm{out}}\right|_{S}$, at two different 
attenuations. Values are given  in units of the SMBH mass $M$
for two different attenuations. The values are listed for various spin parameters
  and pulsar-SMBH distance.}
\end{table}

For
our central SMBH with a mass close to
4$\times$10$^{6}M_{\odot}$ \citep{gillessen2009,ghez2008}, these
differences are on the order of 1 second. The pulsars close to the
Galactic center are expected to be spinning at rates of order 1 second,
thus the spin effects would induce a phase shift on the order of one cycle. These effects, furthermore, originate in the portion of the
photon orbit close to the SMBH, so this order of difference is
maintained even for larger, more relevant values of pulsar-SMBH
distance. The conclusion, based on our limited analysis, is that
pulsar timing of deflected pulses detected at the Earth would be able
to extract the spin signatures of the Sgr A* SMBH.

This estimate is of course for an oversimplified picture. To get
closer to an accurate prediction, we must consider more general
models. A particularly important part of such an analysis would be an
effect that, like spin, induces different delays for prograde and
retrograde beams: the eccentricity of the orbit. Preliminary
calculations suggest that the pattern of timing effects due to
eccentricity of pulsar orbit is, in fact, distinct from the pattern
introduced by the SMBH spin.

\section{Conclusion}
\label{sec:conclusion}
The detection and analysis of pulsar beams strongly deflected by the 
SMBH in Sgr A* holds promise of providing a tool for probing 
strongly curved regions of spacetime. An important question has been 
the probability that such beams are detectable. 
To remedy a shortcoming of  earlier estimates of such probability, we
have considered here
whether the spin of the Sgr A* SMBH can have a significant effect 
on detectibility. We have found that even for the conditions in which 
spin should have its greatest effects
(equatorial photon trajectories, near extreme rotation) the effect of spin
on detectibility is negligible. 

While the effect on detectibility is negligible, we have demonstrated
that the effect of spin on pulsar arrival times is significant enough
that these beams, if detected, can provide us a way of determining the
spin of the SMBH. The fact that this determination is well within the
precision of pulsar timing suggests that the timing can also determine
whether the spacetime near the Galactic center deviates from the Kerr
geometry.  The motivation for an observational program therefore remains 
strong.

We gratefully acknowledge support by the National Science Foundation
under grants AST0545837 and HRD0734800. We also thank the Center for
Gravitational Wave Astronomy, the Center for Advanced Radio Astronomy,
and the Department of Physics and Astronomy at the University of Texas
at Brownsville.

\appendix
\section*{Appendix: Constants of motion from geometrical parameters}
\label{sec:appendix}
We work in a spherical coordinate system centered at the black
hole. As argued before, the pulsars are far enough from the black hole
for us to  assume the spacetime is flat there. From
Figure~\ref{fig:angledefKerr} we then have
\begin{equation}
  \label{eq:defs}
   \vec{s} = r\cos{\theta} \hat{z} + r\sin{\theta} \cos{\phi} \hat{x} + r\sin{\theta} \sin{\phi} \hat{y}\,,
\end{equation}
from which we find
\begin{equation}
   \begin{split}
     \vec{ds}_i  = {} & (\cos{\theta_i}\, dr - r_i\sin{\theta_i} \, d\theta)\hat{z} \\
                    & + (\sin{\theta_i} \cos{\phi_i}\,  dr + r\cos{\theta_i} \cos{\phi_i}\, d\theta - r_i\sin{\theta_i} \sin{\phi_i}\,  d\phi)\hat{x}\\
                    & + (\sin{\theta_i} \sin{\phi_i}\, dr + r_i\cos{\theta_i} \sin{\phi_i} \, d\theta + r_i\sin{\theta_i} \cos{\phi}\,  d\phi)\hat{y}
   \end{split}
\end{equation}
and
\begin{equation}
  \begin{split}
    \vec{p}(\vec{ds}_i)  = {} & \cos{\theta_i}dr \hat{z} + (\sin{\theta_i} \cos{\phi_i} 
dr - r_i\sin{\theta_i} \sin{\phi_i} d\phi) \hat{x} \\
                                  & + (\sin{\theta_i} \sin{\phi_i} dr + r_i\sin{\theta_i} \cos{\phi_i} d\phi) \hat{y}
  \end{split}
\end{equation}
where $r_i, \theta_i$ and $\phi_i$ are coordinates of the photon at the
time of emission in a spherical coordinate system centered at the
black hole. The initial choice of $\phi$ is arbitrary; we set $\phi_i$=0
for simplicity. The above equations  and Equations \eqref{eq:defgammapsi}
give us
\begin{equation}
\label{CalGamma}
     \cos{\gamma}  = \frac{\vec{s}_i\cdot\vec{p}(\vec{ds}_i)}{\lvert 
\vec{s}_i \rvert \lvert \vec{p}(\vec{ds}_i) \rvert} = \frac{dr}{\sqrt{(dr)^2 + (r_i\sin{\theta_i} d\phi)^2}}
  \end{equation}    
so that
  \begin{equation}
    \tan{\gamma}  = r_i\sin{\theta_i} \left(\frac{d\phi}{dr}\right)_i\,.
  \end{equation}
From those equations we also have
\begin{displaymath}
    \cos{\psi}  = \frac{\vec{ds}_i\cdot\vec{p}(\vec{ds}_i)}{\lvert \vec{ds}_i 
\rvert \lvert \vec{p}(\vec{ds}_i) \rvert} 
            \end{displaymath}
\begin{equation}
  = \frac{(dr)^2 + (r_i \sin{\theta_i} d\phi)^2}{\sqrt{(dr)^2 
+ (r_id\theta)^2 + (r_i\sin{\theta} d\phi)^2}\sqrt{(dr)^2 + (r_i\sin{\theta_i} d\phi)^2}},
\end{equation}
from which it follows that   
\begin{equation}
     \tan{\psi}  = \frac{r_id\theta}{\sqrt{(dr)^2 + (r_i\sin{\theta_i} d\phi)^2}} 
= \frac{r_id\theta}{\sqrt{(dr)^2 + (\tan{\gamma} dr)^2}} 
  \end{equation}
  and finally
     \begin{equation}
\tan{\psi}  = r_i\cos{\gamma}\, \left(\frac{d\theta}{dr}\right)_i.       
     \end{equation}

By specifying $\gamma$, $\psi$ and initial values of $r$ and $\theta$,
we obtain initial values of ${d\phi}/{dr}$ and
${d\theta}/{dr}$ which we then use to solve
Equations~\eqref{eq:geodesic} and find the constants of motion.

\end{document}